\documentclass[fleqn,usenatbib]{mnras}

\usepackage{newtxtext,newtxmath}

\usepackage[T1]{fontenc}

\DeclareRobustCommand{\VAN}[3]{#2}
\let\VANthebibliography\thebibliography
\def\thebibliography{\DeclareRobustCommand{\VAN}[3]{##3}\VANthebibliography}

\usepackage{graphicx}	
\usepackage{amsmath}	
\usepackage{amssymb}	

\newcommand{\chandra}{\textit{Chandra}}
\newcommand{\nustar}{\textit{NuSTAR}}

\newcommand{\xmm}{\textit{XMM-Newton}}

\newcommand{\Lya}{Ly$\alpha$}
\newcommand{\kms}{km~s$^{-1}$}

\title[A multiphase UFO in IRAS~13349+2438]{Detection of a possible multiphase ultra-fast outflow in IRAS~13349+2438 with \nustar\ and \xmm}

\author[M. L. Parker et al.]{M. L. Parker,$^{1}$\thanks{E-mail: mparker@sciops.esa.int}
G. A. Matzeu,$^{1}$
W. N. Alston,$^{2}$
A C. Fabian,$^{2}$
A. Lobban,$^{1}$\newauthor
G. Miniutti,$^{4}$
C. Pinto,$^{5}$
M. Santos-Lle\'{o},$^{1}$
and N. Schartel$^{1}$
\\
$^{1}$European Space Agency (ESA), European Space Astronomy Centre (ESAC), E-28691 Villanueva de la Ca\~{n}ada, Madrid, Spain\\
$^{2}$Institute of Astronomy, Madingley Road, Cambridge, CB3 0HA, UK\\
$^{3}$European Space Agency, European Space Research \& Technology Centre (ESTEC), Postbus 299, 2200 AG Noordwijk, The\\ Netherlands\\
$^{4}$Centro de Astrobiolog\'{i}a (CSIC-INTA), Dep. de Astrofis\'{i}ca, E-28691 Villanueva de la Ca\~{n}ada, Madrid, Spain\\
$^{5}$INAF/IASF-Palermo, Via Ugo La Malfa 153, 90146 Palermo, Italy 
}

\date{Accepted XXX. Received YYY; in original form ZZZ}

\pubyear{2020}

\begin{document}
\label{firstpage}
\pagerange{\pageref{firstpage}--\pageref{lastpage}}
\maketitle

\begin{abstract}
We present joint \nustar\ and \xmm\ observations of the bright, variable quasar IRAS~13349+2438. This combined dataset shows two clear iron absorption lines at 8 and 9 keV, which are most likely associated with two layers of mildly relativistic blueshifted absorption, with velocities of $\sim0.14c$ and $\sim0.27c$. 
We also find strong evidence for a series of Ly$\alpha$ absorption lines at intermediate energies in a stacked \xmm\ EPIC-pn spectrum, at the same blueshift as the lower velocity iron feature. This is consistent with a scenario where an outflowing wind is radially stratified, so faster, higher ionization material is observed closer to the black hole, and cooler, slower material is seen from streamlines at larger radii.
\end{abstract}




\section{Introduction}
Since the launch of \chandra\ and \xmm\ in 1999, astronomers have observed blueshifted absorption lines from highly ionized iron in the 7--10 keV energy range \citep[e.g.][]{Pounds03, Chartas02} of active galactic nuclei (AGN) X-ray spectra. These are usually interpreted as evidence for mildly relativistic (0.1--0.3$c$) winds, launched from the black hole accretion disk before crossing the line of sight to the X-ray source. 

Because we can only see the effect of the absorption along the line of sight, it is hard to establish the structure of the absorbing gas. Typically, we only observe absorption lines from a single ionization state and velocity, often only a single Fe~\textsc{xxv/xxvi} line. However, if we can detect more ionization states and velocities for a given UFO, we can start to infer some of the structure of the gas.
To date, only a small number of sources have robust (i.e. confirmed with multiple instruments, including \nustar ) simultaneous detections of multiple ionization states of blueshifted absorption, notably PDS~456 \citep{Reeves18_pds456} and MCG-03-58-007 \citep{Braito18, Matzeu19}. The prevailing interpretation of these dual line detections is that they represent different streamlines in an outflowing wind, with more ionized, faster material coming from smaller radii. 

IRAS~13349+2438 is a $10^9M_\odot$ low redshift \citep[$z=0.10853$][]{Lee13} quasar, which shows a high degree of X-ray variability \citep{Ponti12}, and two layers of warm absorption, outflowing with velocities of $\sim600$~\kms\ \citep{Sako01}. In 2018, using stacked archival \xmm\ spectra of this source, we detected blueshifted lines of Si, S and Fe, with a velocity of $-0.13c$ \citep{Parker18_iras13349}. We also saw weak, possibly background-dependent evidence for a second, more blueshifted Fe absorption line, but we were unable to confirm the detection.
In this letter, we present new observations of IRAS~13349+2438 with \xmm\ and \nustar, focusing on the detection of blueshifted absorption lines in the EPIC-pn and NuSTAR spectra.

\section{Observations and Data Reduction}
We were awarded a 200~ks \nustar\ exposure, simultaneous with a 100~ks \xmm\ exposure, with a primary goal of determining whether the faster absorption feature was genuine. The combination of these two observatories is uniquely powerful detecting UFO lines: combining the high sensitivity and resolution of the EPIC instruments with the high-energy coverage of \nustar , which gives reliable measurements of the continuum shape.

The \nustar\ exposures were split into two: the first 50~ks was taken simultaneously with the \xmm\ observation, and the remaining 150~ks taken two weeks later. There is little variability between the \nustar\ spectra of the two exposures (the 10--50 keV fluxes are $1.5\times10^{-12}$ and $1.6\times10^{-12}$ erg~cm$^{-2}$~s$^{-1}$, respectively), likely because of the high black hole mass and because AGN are generally less variable at high energies.

\begin{table}
    \centering
    \caption{Details of the observations used in this work.}
    \label{tab:obsids}
    \begin{tabular}{l c c c r}
    \hline
    \hline
         Obs. ID & Start Date & Instrument & Clean exposure (ks)\\
    \hline
         60501019002 & 2020-01-20 & FPMA & 50.2\\
                     &            & FPMB & 49.8\\
         60501019004 & 2020-02-03 & FPMA & 147.8\\
                     &            & FPMB & 146.7\\
         0852390101  & 2020-01-20 & EPIC-pn & 73.2$^1$\\
    \hline
    \end{tabular}
$^1$The EPIC-pn was operated in small-window mode, so the live time of the detector was 71\%.
\end{table}

\subsection{NuSTAR}
The \nustar\ data were reduced using the \nustar\ Data Analysis Software (NuSTARDAS) version 1.9.2 and CALDB version 20200510. We extracted source photons from a 30$^\prime\prime$ circular region centered on the source coordinates, and background photons from a larger $\sim100^\prime\prime$ circular region on the same chip, avoiding contaminating sources, the chip gap, and the wings of the source PSF. We bin the FPMA and FPMB spectra to oversample the instrumental resolution by a factor of 3, and to a minimum signal to noise of 6 after background subtraction. The spectra from the two epochs are extremely similar, so we use a time-averaged \nustar\ spectrum for the remainder of the letter.
For plotting purposes, we group the FPMA and FPMB spectra, but they are fit separately in all cases.

\subsection{XMM-Newton}
We reduced the \xmm\ data using the Science Analysis Software (SAS) version 18.0.0 and the latest current calibration files (CCFs) at the time of writing. We extracted source photons from a 20$^\prime\prime$ circular region centered on the source coordinates, and background photons from a larger circular region on the same chip. The EPIC-MOS detectors have higher background contamination at high energies, so are not useful for detecting the potential second Fe feature \citep[the primary UFO features have already been confirmed with the MOS spectra:][]{Parker18_iras13349}. We therefore restrict our analysis to the high signal EPIC-pn data.

The EPIC-pn was operated in small window mode, which excludes the regions of the chip with high copper contamination which can cause spurious features around 8~keV. We binned the spectra to a signal to noise ratio of 6, and to oversample the resolution by a factor of 3.

We also construct a stacked EPIC-pn spectrum with the \textsc{addspec} tool, using all the available archival \xmm\ data as well as the new observation. At high energies (>7 keV) this spectrum is not reliable and shows sharp features below the resolution limit of the detector. This is likely due to combining spectra from different epochs with different levels of background contamination and instrumental gain shift. We therefore restrict our analysis of this stacked spectrum to the lower energy band, where these effects are minimal.

\section{Results}
In this work, we focus on the detection of UFO lines and defer detailed modelling of the continuum and warm absorbers for future analysis. For this reason, we only consider the spectra above 1 keV, and do not include the RGS data which is dominated by the effect of warm absorption. We have undertaken a preliminary inspection of the RGS spectrum to confirm that the absorption is close enough to that found in the archival data that we can safely stack the spectra together and assume the same warm absorption model in our fitting.

We find some moderate variability in the new observations, however a preliminary analysis shows minimal change in spectral shape associated with this variability, so for this work we use the time-averaged spectra. A full analysis of the variability properties will be presented in future work.

\subsection{High energy spectrum}

\begin{figure*}
    \centering
    \includegraphics[width=12cm]{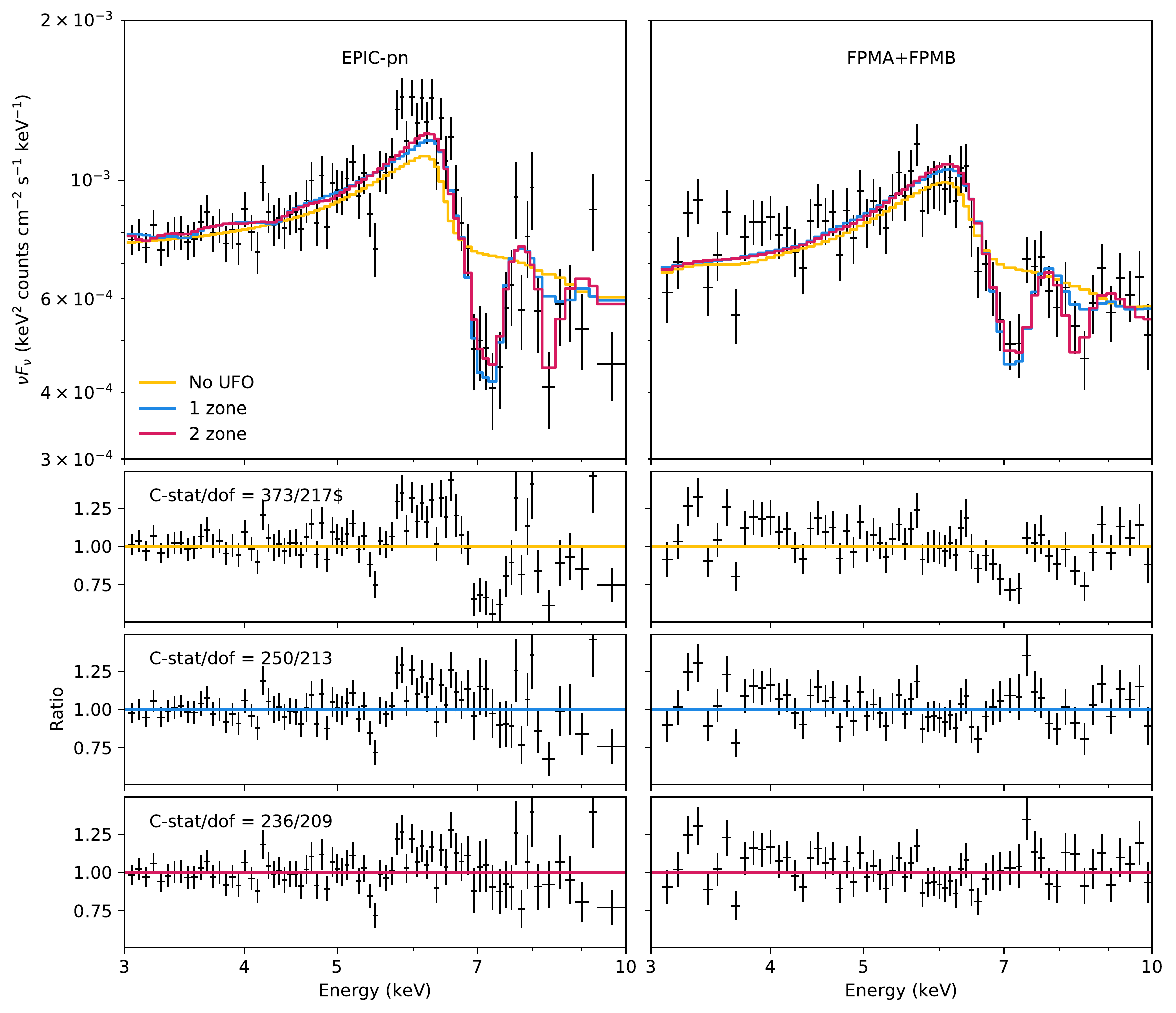}
    \caption{Top panels: Time-averaged EPIC-pn and \nustar\ spectra from the 2020 observations, corrected for effective area but not unfolded from the instrumental response. The \nustar\ spectra are fit from 3--79~keV, but for clarity we only show the 3--10 keV band. The three coloured lines show models fit jointly to the two spectra, with 0, 1 and 2 zones of absorption. Bottom panels: residuals to these three fits.}
    \label{fig:3to10}
\end{figure*}

A simple inspection of the 2020 \xmm\ EPIC-pn and \nustar\ FPMA/B data shows a clear pair of Fe absorption lines at $\sim8$ and $\sim9$~keV (Fig.~\ref{fig:3to10}, top). These correspond to the UFO we detected in \citet{Parker18_iras13349} and the possible secondary feature that we discussed but were unable to confirm. As well as the average \nustar\ spectrum, the line is present in both \nustar\ exposures. The simultaneous detection of this feature with three instruments and in two different epochs leaves little ambiguity - it cannot be due to the XMM background, or noise. 
This feature must be either be due to a second layer of absorption, distinct in ionization and velocity from the primary absorber, or it is a higher order line from the primary absorber.

There is also a clear broad emission feature around 6--7~keV. In \citet{Parker18_iras13349} we investigated the origin of this feature, comparing relativistic reflection with a P-Cygni profile. The reflection model gave a significantly better fit ($\Delta\chi^2=58$, for 2 additional degrees of freedom), and a fit with both models combined implied that the majority of the line emission is due to disk reflection. However, the model used for the P-Cygni profile is a relatively simple phenomenological model, so it is possible that a more sophisticated model would give a better fit. We will explore this in more detail in a follow-up paper (Matzeu et al., in prep). For this work, we focus on the reflection interpretation, which gives a superior fit to the continuum.

Fitting the joint \nustar\ and \xmm\ spectrum above 3~keV with a relativistic reflection continuum, modelled with \texttt{relxilllp} \citep{Garcia14}, and two Gaussian lines gives energies of $7.91\pm0.05$~keV and $9.26\pm0.06$~keV for the two lines, consistent with the ratio of 1.18 between the Fe\textsc{xxvi} \Lya\ and Ly$\beta$ lines. However, the equivalent width of the higher energy line is very high for the Ly$\beta$, 2/3 the equivalent width of the \Lya\ line. We therefore consider it more likely that this represents an separate layer of absorption, and is probably due to highly ionized material that only produces Fe\textsc{xxvi} absorption.
The lines themselves are highly significant: the improvement in fit for adding Gaussians to fit the 8 and 9 keV features is $\Delta$C-stat$=112$ and $\Delta$C-stat$=40$, respectively.


We next replace the two Gaussian absorption lines with two layers of highly ionized absorption, using the \texttt{xabs} model from \textsc{spex} \citep{Kaastra96, Steenbrugge03}, converted to an \textsc{xspec} table model\footnote{available from \url{www.michaelparker.space/xspec-models}.} \citep[for the procedure used to generate this model, see appendix of][]{Parker19_mrk335}. The \texttt{xabs} tables used here assume the default \textsc{spex} ionization curves, from \citet{Steenbrugge05}, and solar abundances. The ionization values returned from this model will be slightly biased by using the assumed SED \citep[for IRAS~13224-3809, the main effect was to bias $\log(\xi)$ by $+0.5$, see][]{Pinto18}
For this fit, we discount the effect of warm absorption, as their impact above 3~keV is negligible. This fit gives a good description of the data (Fig.~\ref{fig:3to10}, bottom panel, table~\ref{tab:highE}), leaving no obvious residuals. Removing the fastest layer and trying to fit the secondary line feature with Fe~K$\beta$ worsens the fit by $\Delta\mathrm{C-stat}=17$, and leaves a line-like residual in both the \nustar\ and \xmm\ spectra.

\begin{table}
    \centering
    \caption{Best-fit parameters for the high energy (>3~keV) spectrum of IRAS~13349+2438, fit with a reflection continuum and two layers of full-covering absorption. All errors are $1\sigma$.}
    \label{tab:highE}
    \begin{tabular}{l c c r}
    \hline
    \hline
        Model & Parameter & Value & Description/unit \\
    \hline
        \texttt{xabs$_1$}   & log($\xi$)        & $4.1\pm0.3$                   & Ionization (erg~cm~s$^{-1}$)\\
                            & $N_\mathrm{H}$    & $0.5^{+5.8}_{-0.2}$           & Column density ($10^{24}$~cm$^{-2}$) \\
                            & $v_\mathrm{rms}$  & $7000\pm2000$                 & 2D RMS velocity (\kms)\\
                            & $v_\mathrm{out}$  & $0.15\pm0.01$                 & Outflow velocity ($c$)\\
        \texttt{xabs$_2$}   & log($\xi$)        & $4.7\pm0.2$                   & Ionization (erg~cm~s$^{-1}$)\\
                            & $N_\mathrm{H}$    & $1.5_{-1.2}^{+8.7}$              & Column density $10^{24}$~cm$^{-2}$ \\
                            & $v_\mathrm{rms}$  & $<8000$                       & 2D RMS velocity (\kms)\\
                            & $v_\mathrm{out}$  & $0.27\pm0.01$                 & Outflow velocity ($c$)\\
        \texttt{relxilllp}  & $a$               & $0.3^{+0.2}_{-0.5}$           & Spin parameter\\
                            & $i$               & $38\pm2$                      & Inclination (degrees)\\
                            & $h$               & $6\pm3$                       & Source height ($r_\mathrm{G}$)\\
                            & $\Gamma$          & $1.9\pm0.1$                   & Photon index\\
                            & $\log(\xi)$       & $3.06_{-0.04}^{+0.11}$        & Ionization (erg~cm~s$^{-1}$)\\
                            & $A_\mathrm{Fe}$   & $5.0_{-0.7}^{+2.6}$           & Iron abundance (solar)\\
                            & $E_\mathrm{cut}$  & $45_{-8}^{+16}$               & Cutoff energy (keV)\\             
                            & norm              & $9_{-4}^{+2}$ &       Normalisation$^1$ ($\times10^{-4}$)\\
    \hline
        C-stat/dof          &                   & 236/209           & Fit statistic\\
    \hline
    \end{tabular}
$^1$See \citet{Dauser16} for the definition of the \texttt{relxill} normalisation.
\end{table}

We next investigate a simple scenario that could result in an enhanced Fe~K$\beta$ line EW, relative to the K$\alpha$ line.
If there are two continua, only one of which is absorbed, a rise in the unabsorbed flux has more effect on the observed equivalent width of a strong absorption line, where the unabsorbed continuum contributes a larger fraction of the flux. This could result in an anomalously strong K$\beta$ line. This could arise naturally in either a wind scenario (by partial covering of the source by clumpy clouds in the wind) or a disk absorption scenario (where only the reflected emission is absorbed). 
We perform a simple test of this model using a partial-covering version of the \textsc{xabs} absorption model, using only a single absorbing layer but allowing the covering fraction to vary. This results in a worse fit ($\Delta$C-stat$\sim15$), and leaves residuals around the secondary line. However, this model is somewhat simplistic and there are complicating factors in both physical scenarios which could result in a better fit, so we cannot fully rule out the second line being Fe~K$\beta$

\subsection{Intermediate energy lines}

To investigate the lower-velocity UFO further we use the stacked EPIC-pn spectrum, constructed from all the \xmm\ exposures, to search for lower energy lines. When fitting this region of the spectrum, we include the two layers of warm absorption identified by \citet{Sako01}, with parameters fixed at the values from \citet{Parker18_iras13349}. We performed a preliminary inspection of the 2020 RGS spectrum to check that the absorption is consistent with that in earlier observations. There are some minor differences, which we will examine in future work, but these are negligible at the resolution of the EPIC-pn.

\begin{figure}
    \centering
    \includegraphics[width=\linewidth]{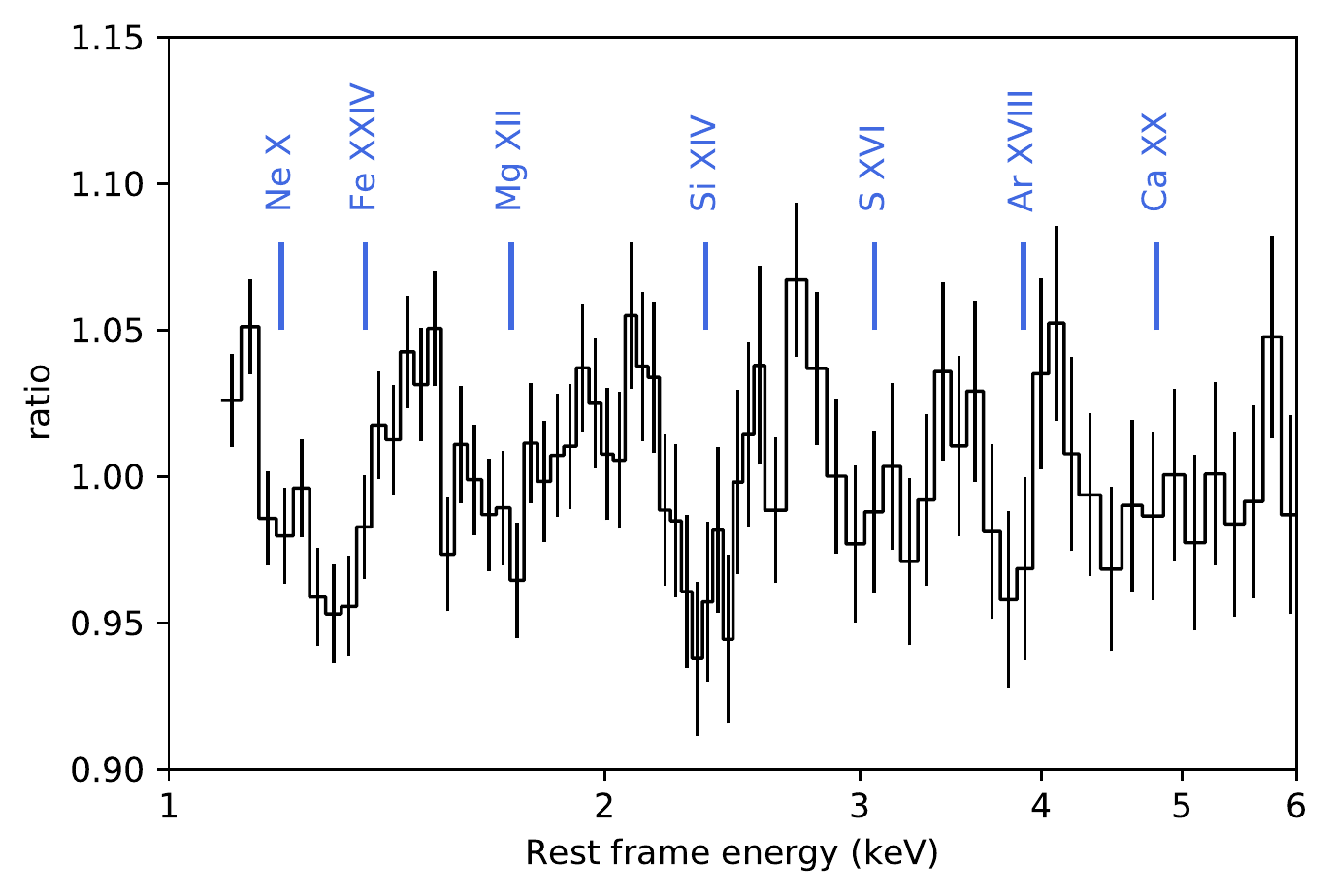}
    \includegraphics[width=\linewidth]{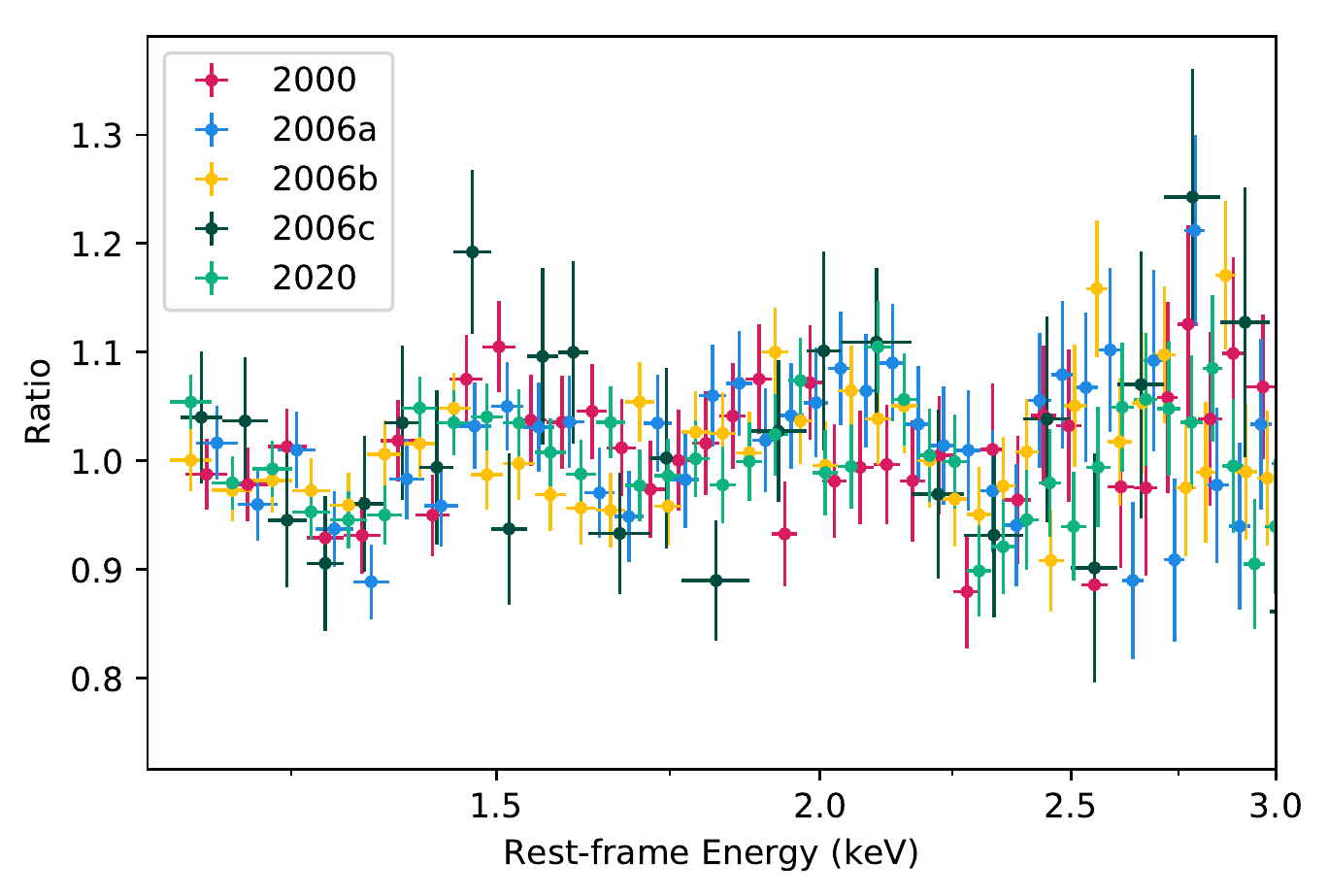}
    \caption{Top: Residuals to a power-law model for the intermediate energy combined EPIC-pn spectrum from all observations of IRAS~13349+2438. Vertical lines indicate the \Lya\ lines blueshifted by a velocity of $0.13c$. Bottom: The same residuals for each observation individually, showing the consistent Ne, Mg, and Si lines. Above 3~keV the data is too noisy and lines cannot be distinguished.}
    \label{fig:intermediatelines}
\end{figure}

We show the residuals of the 1--6~keV spectrum, fit with a phenomenological double power-law plus warm absorption, in Fig.~\ref{fig:intermediatelines}, with the strongest high-ionization absorption lines at a velocity of -0.13~c labelled. There is a clear pattern of broad dips, corresponding to the \Lya\ lines of Ne to Ca and the Fe\textsc{xxiv} triplet (rest frame energy $\sim1.1$~keV).

Individually, the statistical improvement from fitting these lines is relatively small (with the notable exception of the strong Si\textsc{xiv} line), but collectively the significance is very high. We test this by adding a single layer of \texttt{xabs} absorption to the power-law continuum. This improves the fit by $\Delta$C-stat$=47$, to C-stat/dof of 131/113, for four additional degrees of freedom, a cumulative detection significance of over 5$\sigma$.

\begin{table}
    \centering
    \caption{Best-fit parameters for UFO absorption applied to a phenomenological double power-law model from 1--6~keV.}
    \label{tab:intermediate}
    \begin{tabular}{l c c r}
    \hline
    \hline
        Model & Parameter & Value & Description/unit \\
    \hline
        \texttt{xabs}       & log($\xi$)        & $3.9\pm0.1$               & Ionization (erg~cm~s$^{-1}$)\\
                            & $N_\mathrm{H}$    & $1.8\pm0.7$               & Column  ($10^{24}$~cm$^{-2}$)\\
                            & $v_\mathrm{rms}$  & $<9000$                   & 2D RMS velocity (\kms)\\
                            & $v_\mathrm{out}$  & $0.132\pm0.006$            & Outflow velocity ($c$)\\
        \texttt{powerlaw$_1$}   & $\Gamma$      & $1.69\pm0.06$             & Photon index\\
                            & norm              & $9.8\pm1$                 & $10^{-4}$ cm$^{-2}$~s$^{-1}$~keV~$^{-1}$\\
        \texttt{powerlaw$_2$}   & $\Gamma$          & $4\pm1$ & Photon index\\
                            & norm              & $4\pm1$         & $10^{-4}$ cm$^{-2}$~s$^{-1}$~keV~$^{-1}$\\
    \hline
        C-stat/dof          &                   & 131/113           & Fit statistic\\
    \hline
    \end{tabular}
\end{table}

To check that these features are not an artefact of the stacking process, we plot the same residuals for each EPIC-pn spectrum individually (Fig.~\ref{fig:intermediatelines}, bottom). While the data are too noisy above 3~keV for lines to be identified, the lower energy lines are clearly present and consistent with each other between observations.

\section{Discussion}

We consider it highly likely that the intermediate energy lines found in the stacked EPIC-pn spectrum and the slower of the two Fe absorption lines found in the high energy \nustar/EPIC-pn spectrum are from the same material.
The parameters of the absorption used to fit the intermediate energy lines are within 1$\sigma$ error of those of the slower absorption found in the high energy spectrum.

\subsection{Origin of the second Fe line}

The second Fe absorption feature, located at a rest-frame energy 9~keV, must be due to either absorption from Fe\textsc{xxvi} in an additional layer, the Fe\textsc{xxvi} K$\beta$ line, associated with the 0.14$c$ absorption, or a combination of the two, which is the scenario implicitly assumed in our double-absorption model. 

We consider a pure K$\beta$ origin unlikely, as the line is very strong relative to the K$\alpha$ line, roughly twice the expected 1:3 ratio \citep{Krause79}. The observed ratio could be modified by the absorption having a covering fraction less than unity, in which case the stronger K$\alpha$ line would drop in equivalent width more than the K$\beta$ line, however we find that a statistically better fit is obtained with two absorption layers than a single layer, even if that layer is allowed to partially cover the source (which can affect the K$\alpha$/K$\beta$ ratio). 

We conclude that a combination of the K$\beta$ line from the first layer of absorption and the K$\alpha$ line from a second layer of absorption is likely the reason for the 9~keV feature, although we cannot be completely sure, as more complex models that we have not considered may be able to produce a stronger K$\beta$ feature.

\subsection{Dynamics}
Assuming that both layers of absorption are due to an outflowing wind, we can make approximate estimates of their relative locations and the amount of power in each layer. The parameters for the slower UFO do not differ significantly from those found in \citet{Parker18_iras13349}, so for that absorption we use the same derived quantities.

A lower limit on the radius of a wind comes from requiring that the wind velocity exceed the escape velocity, i.e. $v>(2GM/r)^{1/2}$. For the faster absorption, this gives a lower limit of $r>4\times10^{15}$~cm ($27r_\mathrm{G}$), compared with $1.3\times10^{16}$~cm ($89r_\mathrm{G}$). 
An upper limit comes from requiring that the flux be high enough to ionize the gas to the observed level, given the column density (Table~\ref{tab:highE}): $r<L_\mathrm{ion}/\xi N_\mathrm{H}$ \citep[taking the ionizing luminosity of $\sim1\times10^{45}$--$\sim1\times10^{46}$~erg~s$^{-1}$ from the X-ray and EUV fluxes from SED modelling in][]{Lee13}. This gives a range of $1.5\times10^{17}$--$1.5\times10^{18}$~cm (1000--10000$r_\mathrm{G}$) for the slower absorber, and $1.3\times10^{16}$--$1.5\times10^{17}$~cm (90--900$r_\mathrm{G}$) for the faster absorber.

Making the conservative assumption that the winds are located at the radii corresponding to the escape velocity and using the formula from \citet{Krongold07}, assuming the factor describing the opening and viewing angles of the wind, $f(\delta,\phi)=1.5$, we find mass outflow rates of $2.8\times10^{26}$~g~s$^{-1}$ and $2.6\times10^{25}$~g~s$^{-1}$ ($\sim4 M_\odot$ per year), respectively, and kinetic powers of $2.8\times10^{45}$~erg~s$^{-1}$ and $8.5\times10^{45}$~erg~s$^{-1}$ (0.02 and 0.07~$L_\mathrm{Edd}$). These values are approximate, but they are consistent with a wind model where multiple streamlines cross the line of sight, with faster, hotter material located at smaller radii \citep[e.g.][]{Fukumura15}, and meet the predicted 0.5--5\% threshold for AGN feedback \citep{DiMatteo05,Hopkins10}

An alternative model for strongly blueshifted absorption lines was presented by \citet{Gallo11_ufos, Gallo13_ufos}, and recently shown to fit the spectrum of IRAS~13224-3809 by \citet{Fabian2020_ufos}. In this scenario, the blueshifted absorption arises from a surface layer on the disk where these relativistic velocities arise naturally. While this is not the focus of this letter, we note that this model could produce an apparently weaker Fe~K$\alpha$ line as only the reflected emission is affected by the absorption, so an increase in the powerlaw continuum would reduce the apparent equivalent width of the stronger K$\alpha$ line more than K$\beta$.



\subsection{Comparison with PDS 456}
Unlike the faster UFO zone identified in PDS~456 \citep{Reeves18_pds456}, this feature appears to be persistent. Assuming that the possible feature identified at the same energy in \citet{Parker18_iras13349} is the same as the feature we find with the new observations, then it has persisted for over a decade without significant changes. The most likely explanation for this is that the \xmm\ and \nustar\ observations of IRAS~13349+2438 have all occured at roughly the same flux level, despite the high variability of the source on long timescales \citep{Parker18_iras13349}. The strength of the UFO lines in PDS~456 appears to be flux dependent \citep{Parker18_pds456}, as in other sources \citep{Parker17_nature, Parker17_irasvariability, Pinto18, Igo20}, and the range of fluxes observed in PDS~456 is much greater than in IRAS~13349+2438 \citep[e.g.][]{Matzeu17_flares}. It therefore seems likely that the equivalent material in PDS~456 is fully ionized for a large fraction of the time. Further observations of IRAS~13349+2438 aimed at high or low flux states should be able to determine whether the same behaviour takes place in this source.

\section{Conclusions}
We have presented new X-ray observations of the quasar IRAS~13349+2438 with \nustar\ and \xmm. The spectra show a firm detection of a second Fe absorption line, likely from a second layer of gas with a velocity of $v\sim0.27c$.
This is consistent with a stratified wind scenario, where faster streamlines, higher ionization streamlines originate from the inner accretion disk, and slower, colder streamlines are launched further out. 

We also find multiple \Lya\ lines at low energies in a stacked \xmm\ EPIC-pn spectrum including all archival data. These lines are consistent with the blueshift of the lower velocity Fe line, suggesting a common origin.

\section*{Acknowledgements}
MLP and GM are supported by ESA research fellowships.
This letter was written during the 2020 Covid-19 outbreak. The authors would like to thank the medical staff of our respective countries for their work fighting the global pandemic. We thank the referee for their detailed and constructive feedback.

\section*{Data availability}
The data underlying this article are subject to an embargo of 12 months from the date of the observations. Once the embargo expires the data will be available from the XMM-Newton science archive (\url{http://nxsa.esac.esa.int/}) and the NuSTAR archive (\url{https://heasarc.gsfc.nasa.gov/docs/nustar/nustar\_archive.html}).




\bibliographystyle{mnras}
\bibliography{iras13349_bibliography} 








\bsp	
\label{lastpage}
\end{document}